# LEADING DEVOPS PRACTICE AND PRINCIPLE ADOPTION

Krikor Maroukian[1] and Stephen R. Gulliver[2]

[1]Microsoft, Kifissias Ave., Athens, Greece
krmaro@imicrosoft.com
[2]University of Reading, Henley Business School, Business Informatics Systems and Accounting, Reading, UK
s.r.gulliver@henley.ac.uk

*ABSTRACT*

*This research, undertaken in highly structured software-intensive organizations, outlines challenges associated to agile, lean and DevOps practices and principles adoption. The approach collected data via a series of thirty (30) interviews, with practitioners from the EMEA region (Czech Republic, Estonia, Italy, Georgia, Greece, The Netherlands, Saudi Arabia, South Africa, UAE, UK), working in nine (9) different industry domains and ten (10) different countries. A set of agile, lean and DevOps practices and principles, which organizations choose to include in their DevOps adoption journeys were identified. The most frequently adopted structured service management practices, contributing to DevOps practice adoption success, indicate that those with software development and operation roles in DevOps-oriented organizations benefit from the existence of highly structured service management approaches such as ITIL®.*

## 1. INTRODUCTION

The software product development industry is increasingly focused in the pursuit to unlock the full potential of its workforce. There is a requirement to deliver value by securely adding more reliable software more quickly. There is immense pressure to support the existing software product portfolio and develop new versions that contain richer features and fewer defects. Therefore, the adaptability of the IT organization to rapidly changing business demand is becoming, in its turn, increasingly important in delivering value to the customer experience. Business demand is translated to frequent releases, powered by automated build, testing and deployment processes whereby automation reduces required effort to setup new product releases. To that extent business demands need be translated to highly daily commitment to code generation, whilst improving quality assurance, enhanced collaboration and communication, improved visibility of implemented features to the customer, and better testing with customers.

In a world where every Company is a software company [1] and software is eating the world [2], adaptability has become the new competitive advantage, shifting commercial focus from position, scale, and "first order" capabilities in producing or delivering an offering, to "second order" organizational capabilities that foster rapid adaptation [3].

Traditional structured approaches in software product development and project management in software intensive industries have had to IT project failure [4] [5] [6] [7] [8] and software project failure [9] [10] - with a number of examples from the public sector [11] [12] [13]. In addition, traditional software security has also witnessed research by academia on this topic

[53]. Considering this, organizations have focused on fostering agility in their software development and operations team structures. These organizational changes entail transitioning software development practices, transition team practices, transitioning management approach, transitioning reflective practices and transitional culture [15].

## 1.1. Structured, Agile, Lean and DevOps Challenges

Structured IT service management (ITSM) frameworks such as ITIL® [16], and project management frameworks such as PRINCE2® [17] and PMBOK® [18] have been introducing numerous decision making roles and gates in IT organizations, and thus have introduced numerous delays in the product development lifecycle. In addition, accountability in structured approaches supports increased culpability in process ownership, which although such tests increase accountability they reduce flexibility, since all changes require the approval of multiple stakeholders. Furthermore, structured approaches to change, release, and deployment management of new products and services within the IT industry, has led to the innate proclivity to be blameful within post implementation reviews, or within post-project delivery lessons-learned meetings.

Agile, lean, and DevOps principles and practices aim to identify value and non-value adding activities within ITSM processes. Specifically, the identification regards the end-to-end ownership of associated roles, processes and technology [19] to the software product development lifecycle [20] [21] [22]. IT organizations aiming to adopt agility find that the more defined processes results in restricted agility [23]. Therefore, there is clearly a need to extend, and or shift, from structured service management practices towards agility and leanness. The transition, i.e. from a framework or process-led organizational environment to the adoption of groups of best practices, entails a significant shift in individual and organizational mindset. There needs to be a clear organization-specific roadmap on the types of practices and principles that need to be adopted, including i) team structures that needs to be applied, and ii) leadership styles that can help guide others towards agility/leanness adoption.

This research firstly aims is to identify the practices and principles that Agile, lean and DevOps communities have developed, in regard to product development and its overlap with ITSM processes. Secondly, to realize the effect that Agile, Lean and DevOps practice and principle adoption has on structured service management processes. Finally, as a consequence, it is important to realize whether Agile, Lean and DevOps practice and principle adoption requires any sort of leadership needs, and whether these leadership needs already form part of an individual leader role or team structure. This aim is reflected in the defined research questions: **RQ1)** Which agile, lean and DevOps practices / principles can improve productivity in a business environment that has adopted a structured service management approach? **RQ2)** Can DevOps-oriented environments benefit from structured service management practices? **RQ3)** Can Leadership affect DevOps adoption within an organization and to which extent?

## 2. DEFINING AGILE, LEAN, AND DEVOPS PRACTICES AND PRINCIPLES

The Fourth Industrial Revolution, characterized by the growing utilization of disruptive digital technologies, is transforming the world of work; e.g. both the jobs and the skills that are needed in business to compete. Moreover, research by McKinsey [24] suggests that globally about half of the jobs performed today by humans will be disrupted in some way by automation, and the World Economic Forum [25] stated that 42% of the core job skills required today are set to substantially change by 2022. In addition, leading cultural change will be key to digital business transformation [26]. Within this dynamically changing business world, use of software management is playing a much larger and more strategic role in shaping how companies compete, with large 'traditional' organizations finding themselves limited in their ability to respond to market and customer needs.

## 2.1. Agile Software Development

During the 1990s, individuals with a desire to think and act outside the structured approaches imposed in project and product management began forming the agile community; a term formally coined in 2001 Agile Manifesto [27]. The manifesto set out to establish principles to improve the existing software development approaches. Agility aimed at solving a lot of the issues that were created in information intensive organizations by structured approaches. In addition, Agile Software Development (ASD), which emerged in 2001 as an evolutionary practice to existing structured approaches, advocated for iterative short-cycled development increments and continuous integration - as opposed to structured engineering stage-gate models [28]. SCRUM [29], i.e "a framework within which people can address complex adaptive problems, while productively and creatively delivering products of the highest possible value." [30], is commonly used as an agile product development approach in software-intensive organizations.

## 2.2. Lean Mindset

The roots of Lean Enterprise stretch as far back as 1908 – i.e. to a time when Henry Ford's Ford Motor Company was designing and producing Ford Model T automotive cars. The grandiose Model T mass production plan was successful because it provided inexpensive transportation, which symbolised both innovation and modernization for the rising middle classes in the US. The set of practices and principles employed by Henry Ford's automotive production factories developed to what is known as Ford Production System (FPS). Moreover, FPS became the baseline synthesis of lean manufacturing [31]. Henry Ford extended organizational considerations to human psychology which aimed at an inclusive work environment where each and every one factory employee partnered with the organization to achieve its goals. Following World War II, FPS was transformed by Toyota into two pillars known as i) Just In Time (JIT) and ii) Jidoka aka autonomation [32] [33] – making kanban boards, kaizen (continuous improvement), and poka-yoke (error-proofing) a key part of the Toyota Production System (TPS) [32] [33].

In the early 20th century Japan was already adopting a lot of the FPS techniques and adapting them to the proven methods for automotive mass production purposes aiming for cost efficiencies and increased quality. Developing a Lean Enterprise is all about eliminating friction and introduced waste in the value stream [59] [60] and reducing the time taken to deliver a product or service to market consumers. The term "Lean" was coined in 1988 by John Krafcik [34] and popularized in 1990 by James P. Womack [35], with the aim to remove the following waste: 1) partially completed work, 2) unneeded product features, 3) relearning/skilling of staff, 4) poor handoff, 5) task switching, 6) delays, 7) product defects [36], and a later addendum 8) underutilized staff. Lean IT's providers aim to transpose the same approaches to waste to software development, i.e. to eliminate or reduce their impact on product development lead times to market delivery. In comparison to ASD, it is notable that Lean Software Development (LSD) was an incremental improvement [37].

## 2.3. DevOps and its Adoption

DevOps offers an unprecedented opportunity for organizations to transform their Software Development lifecycle to increase efficiency and meet end-users' changing expectations. DevOps attempts to redefine the foundations of software development and management recasting the approach concerning development of every element [38] even in cloud services provisioning [14]. The reformation that DevOps brings, with its set of developed practices, also extends to the customer experience.

There are a number of terms and variety of practices and definitions that software practitioners use when defining DevOps [39] [40] [41] [42] [43] [44]. In practice the use of different DevOps definitions leads to unnecessary confusion when it comes to IT organizations adopting a 'DevOps-oriented mindset'. Moreover, the numerous associated acronyms that accompany DevOps has a significant role to play in the result of indecisiveness or definition diversity. DevSecOps [45] or SecDevOps (Development-Security-Operations), BizDevOps (Business-Development-Operations) [46] and DevNetOps [21], are all part of the DevOps definition held within organizations. The majority of the descriptions specify DevOps as a term that is used to emphasize the collaboration between software development and operations. Additionally, there is a growing requirement from the research industrial communities to define DevOps [43]. There is also published research work that downplays the fact of not having consensus over a DevOps definition [21].

However, DevOps is more than just a mindset but rather patterns of DevOps practices [41]. In Agile software development there is a distinction between practices and influences [47] which can be extended by a lean principles background that form a prerequisite for successful DevOps adoption [44]. Furthermore, there is research that categorizes advisory skills, testing skills, analysis skills, functional skills, social skills, decision making skills and full stack development skills as the skillset that can result to successful DevOps cross-functional teams [48]. This can be further complemented by a set of practices (common among development and operations teams, development-specific, operations-specific) and a set of principles (social aspects, automation, quality assurance, leanness, sharing measurement) [40]. This is closely linked to CAMS (Culture-Automation-Measurement-Sharing) model originally coined by John Willis and Damon Edwards [19] and later refined to CALMS (Culture-Automation-Lean-Measurement-Sharing) by Jez Humble. CALMS shares similarities with another model that involves a specific set of categories namely: agility, automation, collaborative culture; also called DevOps Culture [49], continuous measurement, quality assurance, resilience, sharing and transparency [50]. This can be further extended to include collaboration in terms of empathy [44], respect, trust, responsibility and incentive alignment and open communication [51]. There are recurring studies to suggest that the lack of a 'collaborative culture' is detrimental to the success of DevOps teams and DevOps practice and principle adoption in an organisation [14] [40] [44] [48] [49] [51].

### 2.4. Leadership styles relevant to DevOps

The are various leadership styles that should be considered when considering DevOps – especially if a highly structured organization is attempting to adopt agile, lean and DevOps practices and principles. A non-exhaustive list of those leadership styles is provided:

- Transactional Leadership [52]
- Transformational Leadership [53]
- Authentic Leadership [54]
- Servant Leadership [55]
- Ad Hoc Leadership [56]

The State of DevOps Report in 2017 discovered a correlation between transformational leadership and organizational performance [57]. Transformational leadership comprises of four dimensions: idealized influence, inspirational motivation, intellectual stimulation, and individualized consideration [52] and was first posited by James McGregor Burns in 1978 [53].

The State of DevOps Report 2017 report conveys that DevOps leaders with a servant leadership mentality inspired better team performance [57]. In fact, Servant Leadership Theory is a mixture of transformational and transactional styles of leadership. In essence, the leader is serving rather than being served and therefore, creates an environment of trust, collaboration and reciprocal service which ultimately leads to higher performance [52]. On the other hand, ad hoc leadership is constituted of three poles (the team, the customer, the management) as opposed to two poles that formulate other leadership styles and its lifecycle is characterized by a leadership style fading and another one becoming prevalent in a software development team setting [56].

## 3. METHOD AND APPROACH

Having distinguished between agile, lean and DevOps practices and principles described in literature, it is now essential to determine whether these views align with industry domain practitioners.

### 3.1. Research Design and Interview Structure

To capture contextually relevant data, semi-structured interviews were conducted with thirty (30) practitioners in companies working within a wide range of countries (Czech Republic, Estonia, Italy, Georgia, Greece, The Netherlands, Saudi Arabia, South Africa, UAE, UK). All interviewees contributed to DevOps adoption processes in their respective companies. Participants were recruited using two approaches: 1) through direct contact at an ITSM / DevOps event in Europe, and 2) via a general call for participation posted on professional social media networks; including Linkedin and IT societies such as IT Service Management Forum (itSMF) and British Computer Society (BCS) – The Chartered Institute for IT. To achieve a heterogeneous perspective, and to increase the wealth of information, practitioners from a variety of organisations were invited and consulted. Although face-to-face interviews were preferred, a number of web interviews were conducted using a range of online tools (Skype for Business and Zoom). Table 2 presents the characteristics of the participants. To maintain anonymity, in conformance with the human ethics guidelines, we refer to the participants as P1–P30 (see Table 2). At the beginning of each interview the interviewee consented to: i) an audio recording being taken, and ii) the transcript being used only in the context of the research. Instructions were clear to state that no names or organisation titles would be discloses as part of this research.

Interviews were conducted between September 2018 and January 2019. The interviews lasted a minimum of 34 min, a maximum of 67 min, and an average of 50 min. Data collection and analysis was aggregated in order to answer the research questions posed at the end of section 2, and were mapped to interview questions (see Table 1). The whole set of interview questions is available at the following URL: **https://tinyurl.com/ybxrcujq**

Table 1. Research to interview questions mapping.

| Research Question | Interview Question |
|---|---|
| Data collection for segmentation purposes | 1, 2 & 3 |
| R1) Which agile, lean and DevOps practices and principles can improve productivity in a business environment that has adopted a structured service management approach? | 4, 5, 6, 7, 8, 9, 10, 11, 12, 15, 16 |
| R2) Can DevOps-oriented environments benefit from structured service management practices. | 13, 14, 15, 20 |
| R3) How does Leadership affect DevOps adoption within an organisation? | 17, 18 , 19, 20 |

Table 2. Interview participant profile. PX means professional experience in years, CN means country of work and CS means company size (Micro - MC < 5, Small < 50, Medium - M < 250, Large > 251) [58].

| P# | Job Title | PX | CN | Domain | CS |
|---|---|---|---|---|---|
| P1 | PMO Director | 14 | Saudi Arabia | Aviation | L |
| P2 | Principal Consultant, IT Service Management | 13 | Italy | IT Consulting Services | L |
| P3 | CIO | 26 | Greece | Insurance | L |
| P4 | Principal Consultant, IT Service Management | 11 | UK | IT Consulting Services | MC |
| P5 | Managing Director, IT Service Management | 32 | UK | IT Consulting Services | S |
| P6 | Smart Systems Manager | 23 | Greece | IT Consulting Services | L |
| P7 | Senior Digital Transformation Technologist & Solution Practice Lead | 30 | UAE | IT Consulting Services | L |
| P8 | Principal Consultant, IT Service Management | 34 | UK | IT Consulting Services | L |
| P9 | Founding Consultant, IT Service Management | 19 | UK | IT Consulting Services | S |
| P10 | Managing Director | 29 | UK | IT Consulting Services | S |
| P11 | Head of Remote Transactions | 16 | Greece | Banking | L |
| P12 | Consultant | 34 | Netherlands | IT Consulting Services | M |
| P13 | Deputy CIO | 22 | Greece | Construction Management | L |
| P14 | Head of Applications | 18 | Greece | Lottery | L |
| P15 | Principal Consultant, IT Service Management | 21 | South Africa | IT Consulting Services | MC |
| P16 | Founding Consultant, IT Service Management | 34 | UK | IT Consulting Services | MC |
| P17 | Managing Director, IT Service Management | 19 | UK | IT Consulting Services | MC |
| P18 | Managing Director and Lead Consultant | 14 | UK | IT Consulting Services | MC |
| P19 | IT Operations Manager | 13 | Greece | Lottery | L |
| P20 | IT Operations Manager | 15 | UK | Government | M |
| P21 | Founding Consultant, IT Service Management | 34 | UK | IT Consulting Services | MC |
| P22 | Assistant General Manager, IT Operations | 28 | Greece | Banking | L |
| P23 | CDO | 13 | Estonia | Government | L |
| P24 | CIO | 20 | Greece | Insurance | L |
| P25 | CIO | 27 | Greece | Aviation | L |
| P26 | Development Team Lead | 11 | Greece | Lottery | L |
| P27 | IT Operations Lead | 12 | Georgia | Government | M |
| P28 | Business Development Director | 18 | Greece | IT Consulting Services | L |
| P29 | Operations and Innovation Lead, IT Services | 11 | Czech Republic | Courier Services | L |
| P30 | CIO | 28 | Greece | Automotive | M |

## 4. EVALUATION AND RESULTS DISCUSSION

The semi-structured interview, see Table 2, consisted of twenty (20) interview questions (Q1-Q20). The first three questions aimed to collect data on interviewee demographics i.e. job role, industry domain, and working country (see Tables 3 - 4 for demographic breakdown). The country of employment for interview participants included Greece (11), UK (10), Saudi Arabia

(2), Czech Republic (1), Estonia (1), Georgia (1), Italy (1), Netherlands (1), South Africa (1), UAE (1), see also Tables 3-4 for demographic breakdown.

Table 3. Job role of interview participants (interviewee count: 30).

| Job Title | No. of Participants |
|---|---|
| Principal Consultant | 9 |
| Managing Director | 4 |
| CIO | 4 |
| Deputy CIO/Assistant General Manager/CDO | 3 |
| IT Operations Manager | 3 |
| PMO Director | 1 |
| Head of Remote Transactions | 1 |
| Smart Systems Manager | 1 |
| Head of Applications | 1 |
| Development Team Lead | 1 |
| Business Development Director | 1 |
| Operations and Innovation Lead | 1 |

Table 4. Job role of interview participants (interviewee count: 30).

| Industry Segmentation | No. of Participants |
|---|---|
| Consulting Services | 14 |
| Aviation | 3 |
| Government | 3 |
| Lottery | 2 |
| Insurance | 2 |
| Finance | 2 |
| Manufacturing | 1 |
| Logistics | 1 |
| ISV | 1 |
| Automotive | 1 |

Fifteen (15) participants were IT consultants and fifteen (15) were employed at customer organisations - characterised as "service providers" according to ITIL® [16], see Fig. 1.

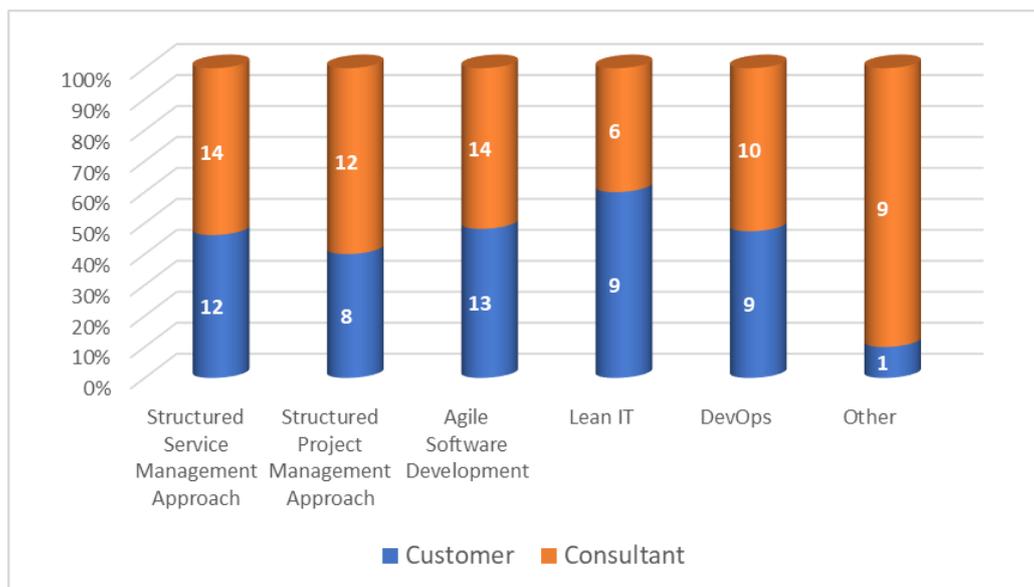

Figure 1. Agile and Lean Practices (Top 3 highlighted) [Interviewee count: 30].

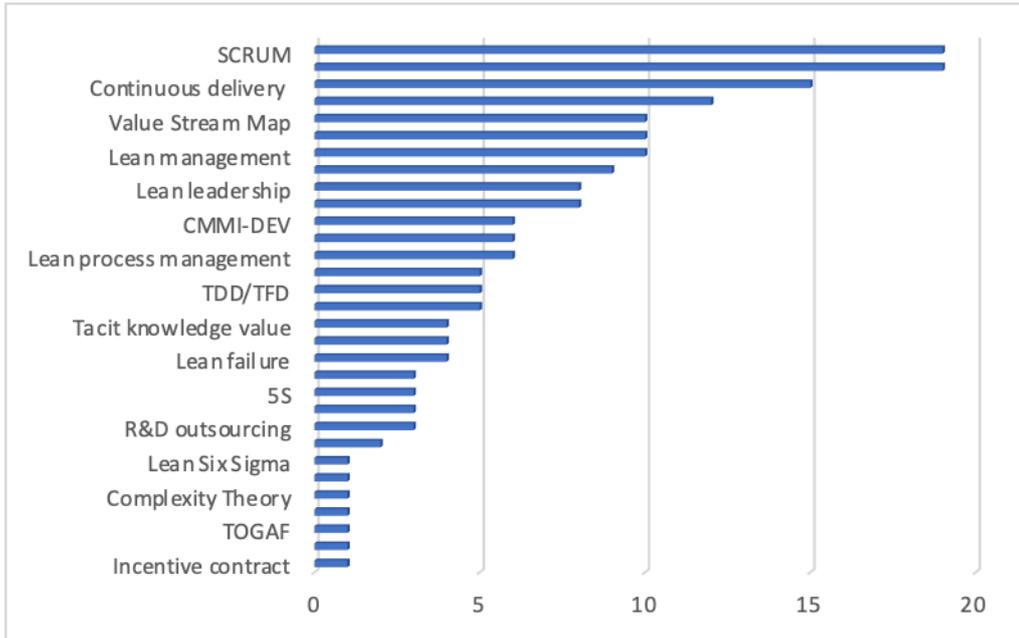

Figure 2. Agile and Lean Practices [Interviewee count: 30].

Interview participants indicated their most preferred structured, agile, and lean practices (see Fig.2) and principles (see Fig.3).

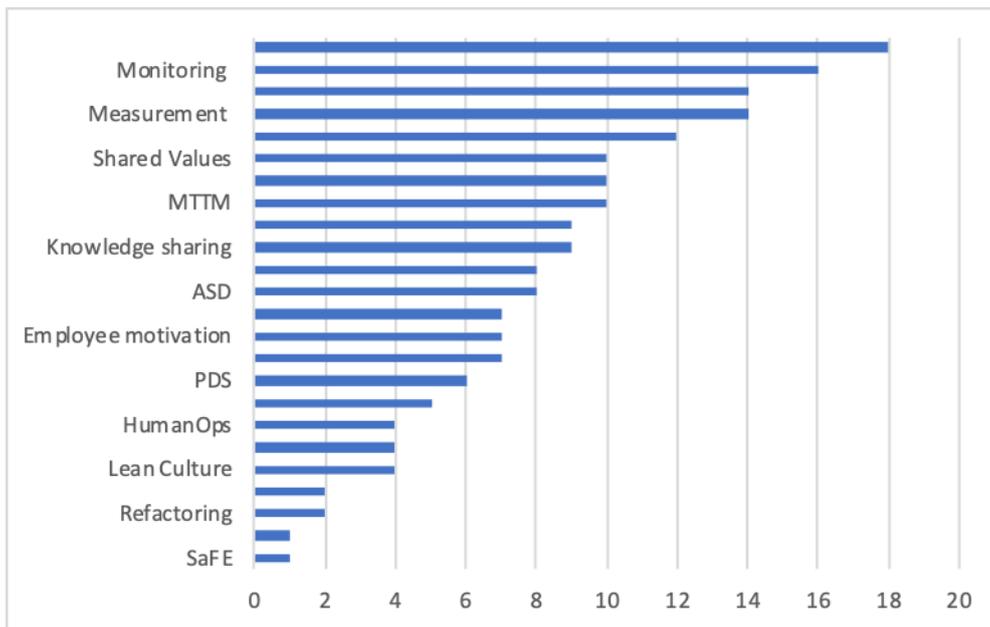

Figure 3. Agile and Lean Principles [Interviewee count: 30].

When considering structured IT service management (ITSM) processes, the interview participants identified a set of practices that contribute to value delivered to software development. Change Management was the most preferred process compared to the rest of the ITSM processes, see Table 5. Additionally, Release and Deployment, Incident and Problem and Service Level Management conclude the top four ITSM processes, which affect value delivery in software development. The prominence of change management was repeated many times throughout the course of interviews with P27 (Georgia, IT Operations Lead) stated that:

> Any change can bring resistance and hinder adoption practices. Moving away from any already established approach generates resistance.

Moreover, P24 (Greece, CIO) added that:

> Resistance happens because all the teams are getting out of their comfort zone. We are talking about different methodology, different structure, different KPIs, different roles, different rewarding scheme, different working location since the team is now collocated - everything is different.

Whereas P20 (UK, IT Operations Manager) states that:

> Change management is not generally well understood within organisations

On the contrary P18 (UK, Managing Director and Lead Consultant) argued that:

> Rather than adopting every new framework, methodology, set of practices, organizations should look into identifying the current bottlenecks and improvement areas.

Table 5. ITSM process significance to value delivery of software development [Interviewee count: 30].

| IT Service Management Process | Adds Value to Software Development (%) |
|---|---|
| Change Management | 24 |
| Release and Deployment Management | 15 |
| Incident and Problem Management | 10 |
| Service Level Management | 9 |
| Availability Management | 7 |

In addition, 66.67% of interviewees agree that agile and lean principle and practice adoption is an extension of established structured ITSM approaches - such as ITIL®. Only 20% stated that a complete replacement of those is required. However, concerns on ITIL adoption were mentioned by P6 (Greece, Smart Systems Manager):

> ITIL is only used for IT operations and too many roles and responsibilities are defined within ITIL, which means that poor adoption leads to increased confusion of the workforce adopting it.

In fact, the extension of principles and practices signals the transition an organisation has to pursue in order to achieve the desired adoption level. However, the top three challenges identified concerning DevOps practice and principle adoption journey were: 1) Poor communication and information flow; 2) Deep-seated company culture; 3) Operations not participating in the requirements specifications. Additionally a number of interviewees registered that blameful culture and time consuming bureaucratic processes do not promote a sense of change in behavior to adopt new practices and principles but maintained a collective cultural complacency among IT teams. P7 (UAE, Senior Digital Transformation Technologist & Solution Practice Lead) mentioned that:

> Blame 'game' exists between IT teams which breads increased blameful culture, especially between Dev and Ops teams. By bringing these two teams together to code, test, deploy - the blame game stops. So now a blame-free culture starts to be promoted and gradually becomes evident as change emerges in behavioral patterns.

P11 (Greece, Head of Remote Transactions) adds that:

> Bureaucratic approach leads to informal ways of complete disregard of approval points. Senior management is keen to use this kind of approach to get things done quicker.

DevOps is highly regarded as a group of practices and principles that characterise collaborative culture [50] and these top three challenges indicate the requirement to address them from an organisational culture perspective. According to answers from Question 4, 66% of participants are aware of DevOps and its associated practices and principles. Therefore, naturally the participants were asked to define DevOps. The four most popular phrases used were "a shift of mindset", "enhanced collaboration and communication", "continuous deployment" and "automated testing process". The shift of mindset was pointing to established organizational cultural behaviors such as the one P3 (Greece, CIO) referred to:

> There is a mindset to "never outshine the master".

P11 (Greece, Head of Remote Transactions) mentioned that:

> The 'email culture' on which business units heavily rely is detrimental to DevOps adoption aspirations.

To that extent P18 (UK, Managing Director and Lead Consultant) mentioned that:

> Culture is a very wide term. So if the incentives are in conflict with team expectations than there is going to be a situation of complaining about tool usage. Enterprise-wide incentives alignment is strongly required under such circumstances.

Moreover, 53% believe that the DevOps leader role should be an individual professional, whereas 33% would trust the role to a team. People suggested that it was best to have an individual lead DevOps adoption, and organisational transformation efforts initially, but that and then transition to a team effort was also deducted at 13%. Note that the adoption efforts should be continuous in nature, and not be conducted in a project-based manner as temporary endeavor. In this context P18 (UK, Managing Director and Lead Consultant) stated that:

> DevOps adoption practices and principles should not be viewed as a project under the context of a transformation with a beginning and an end rather a continuous aspiration for improvement of the current state of adopted practices and principles.

In addition, P8 (UK, Principal Consultant, IT Service Management) added that a common pitfall is that:

> Overestimation of DevOps practice adoption is common.

P10 (UK, Managing Director) mentioned one area that requires particular attention:

> Uneven experience of people gives birth to assumptions. For instance, if not everyone in the same team has the same level of knowledge and understanding on ITIL then different people would assume different definition for IT service management. HR plays a big role in recruiting people with uneven skills. This is an unrecognised cost to the IT organization.

Furthermore, P21 (UK, Founding Consultant, IT Service Management) stated that:

> The transformation of Waterfall-to-Agile-to-DevOps in an IT organization has to be an enterprise-wide endeavor. The missing link is HR not being on the same page with the efforts to change towards agility.

P1 (Saudi Arabia, PMO Director) added that:

> The human resources department is an enabler leading the change.

Whereas P14 (Greece, Head of Applications) commented that:

> Lack of continuous commitment to DevOps adoption by organization-internal IT customers inhibits the adoption itself.

The leadership skills categories that were mentioned by > 50% of interview participants included: 1) technical background; 2) negotiation skills; 3) communication and collaboration skills; and 4) previous experience on transformation. Holistic systems thinking was mentioned by 27% of interviewees. Business background by 17%. Strategic thinking by 13%. Furthermore, there was a lot of iteration around the influential skills, holistic systems thinking, a multi-cultural mindset and increased awareness around dealing with suboptimal productivity.

When considering DevOps leadership objectives, a remarkable 87% of interview participants agreed that DevOps practice adoption should be extended in an enterprise-wide fashion and should include external service providers in its scope. To overcome DevOps adoption inhibitors P19 (Greece, IT Operations Manager) stated that:

> Leadership skillset is the most important thing to adoption barrier breakdown.

In addition, P23 (Estonia, CDO) added that:

> A cross-functional leadership role with end-to-end ownership of DevOps adoption is imperative.

Lastly, the organizational teams should be part of a DevOps practice adoption journey are IT Development (97%), IT Operations (97%), Quality Assurance (93%), Information Security (80%) and Board of Directors (73%).

## 5. THREATS TO VALIDITY

Concerning construct validity, there is heavy reliance on each of the interviewed practitioners' subjective perception. However, currently there is no objective approach to measure whether or not a DevOps transition journey, in the context, of practice and principle adoption within organizations can be associated to successful outcomes. The semi-structured interview series approach undertaken offers rigorous procedures for data analysis but with a certain degree of research bias. It is probable, that other researchers might deduce different findings and outcomes looking at the same set of data, but the author believes the main perceptions would be preserved. This is a typical threat related to similar studies, which do not claim to generate definitive findings.

The author welcomes extensions to the research or potential discovery of new dimensions for future study. Future work can focus on the identification of DevOps adoption leadership styles or leader characteristics that could "make" or "break" a transition journey towards a DevOps-oriented organization. Furthermore, concerning external validity, although the viewpoint of the interviewed practitioners is considered with different backgrounds, working in organizations from nine (9) different industry domains and ten (10) different countries the author does not claim that research results from this contribution are valid to other scenarios. However, saturation was achieved after the 20th interview.

## 6. FUTURE RESEARCH

This study can be further enhanced in the future by assessing and determining the usefulness of the outcomes under the prism of a survey which reiterates the questions posed to a wider participation poopulation. The extension of the findings can be further evaluated under the lens of a case study.

# 7. CONCLUSIONS

The data collected from a series of interviews and participating practitioners, indicate a clear list of specific agile, lean and DevOps practices and principles that are regarded as an extension to structured service management approaches, and are relevant to understanding DevOps adoption. The main findings associated to the research questions are that:

1. Specific agile, lean and DevOps practices such as 1) organizational culture, 2) monitoring/measurement, 3) automation are crucial in the software development lifecycle (RQ1)

2. Specific agile, lean and DevOps principles such as 1) SCRUM 2), Kanban 3) Continuous Delivery are crucial in the software development lifecycle (RQ1)

3. The set of service management processes that continue to form a strong part of DevOps-oriented structures are Change Management, Service Portfolio Management (including Service Catalog Management), Release and Deployment Management and Service Level Management. (RQ2)

4. There is overwhelming consensus that a DevOps leadership role should exist (86%) and that the role should carry a continuous effect not a project based. (RQ3)

5. DevOps practices and principles adoption are challenged due to poor communication and information flow, deep-seated company culture and operations not being involved in the requirements specifications. (RQ3)

6. DevOps practice adoption should be extended in an enterprise-wide fashion (87%), with team structure based on existing Development (97%), Operations (97%), Quality Assurance (93%) and Information Security (80%) teams. (RQ3)

The outcomes of this paper can be used by practitioners in software-intensive organisations willing to introduce a DevOps orientation in terms of practices and principles adoption. The research can further be extended in the future to explore more of the facets of leadership style(s), capabilities, skills and competencies required in the context of continuous DevOps adoption.

## ACKNOWLEDGEMENTS

The authors would like to thank the interview participants for all the time and committed efforts to contribute to the results and outcomes towards research questions.

## REFERENCES

[1] Computer Weekly Digital Edition, https://www.computerweekly.com/news/2240242478/Satya-Nadella-Every-business-will-be-a-software-business. Accessed April 18, 2020

[2] The Wall Street Journal, https://www.wsj.com/articles/SB10001424053111903480904576512250915629460. Accessed April 18, 2020

[3] Reeves, M. Deimler, (2011) "Adaptability: The New Competitive Advantage. In: Harvard Business Review", *Harvard Business Review Press Books*, pp1-9.

[4] Lauesen, S. (2020) "IT Project Failures, Causes and Cures," *in IEEE Access*, Vol.8, pp72059-72067, 2020.

[5] Diaz Piraquive, F. N., Gonzalez Crespo, R., Medina Garcia, V. H. (2015) "Failure cases in IT project Management," *in IEEE Latin America Transactions*, Vol.13-7, pp2366-2371.

[6] Gupta, S., Mishra, A., Chawla, M. (2016) "Analysis and recommendation of common fault and failure in software development systems," *2016 International Conference on Signal Processing, Communication, Power and Embedded System (SCOPES)*, pp1730-1734.

[7] Liu, S., Wu, B., Meng, Q. (2012) "Critical affecting factors of IT project management," *2012 International Conference on Information Management, Innovation Management and Industrial Engineering*, pp494-497.

[8] Altahtooh, U. A., Emsley, M. W. (2014) "Is a Challenged Project One of the Final Outcomes for an IT Project?," *47th Hawaii International Conference on System Sciences*, pp4296-4304.

[9] Verner, J., Sampson, J., Cerpa, N. (2008) "What factors lead to software project failure?," *2nd International Conference on Research Challenges in Information Science*, pp71-80.

[10] Sardjono, W., Retnowardhani, A. (2019) "Analysis of Failure Factors in Information Systems Project for Software Implementation at The organization," *International Conference on Information Management and Technology (ICIMTech)*, Jakarta/Bali, pp141-145.

[11] Ashraf, J., Khattak, N.S., Zaidi, A.M. (2010) "Why do public sector IT projects fail," *7th International Conference on Informatics and Systems (INFOS)*, pp1-6.

[12] Abbas, A., Faiz, A., Fatima, A., Avdic, A. (2017) "Reasons for the failure of government IT projects in Pakistan: A contemporary study," *International Conference on Service Systems and Service Management*, pp1-6.

[13] Mohagheghi, P., Jørgensen, M. (2017) "What Contributes to the Success of IT Projects? Success Factors, Challenges and Lessons Learned from an Empirical Study of Software Projects in the Norwegian Public Sector," *IEEE/ACM 39th International Conference on Software Engineering Companion (ICSE-C)*, pp371-373.

[14] Rajkumar, M., Pole, A.K., Adige, V.S., Mahanta, P. (2016) "DevOps culture and its impact on cloud delivery and software development," *International Conference on Advances in Computing, Communication, & Automation (ICACCA)*, pp1-6.

[15] Hoda, R., Noble, J. (2017) "Becoming Agile: A Grounded Theory of Agile Transitions in Practice," *IEEE/ACM 39th International Conference on Software Engineering (ICSE)*, pp141-151.

[16] Office of Government Commerce (OGC) (2011) "ITILv3 Core Publications", *TSO (The Stationery Office)*, UK.

[17] Office of Government Commerce (OGC) (2017) "Managing Successful Projects with PRINCE2", *TSO (The Stationery Office)*, UK.

[18] Project Management Institute (PMI) (2017) "A guide to the project management body of knowledge (PMBOK® Guide), 6th ed.", *Project Management Institute*, Inc, Pennsylvania, USA.

[19] Willis, J., 2010. What DevOps means to me, https://blog.chef.io/what-devops-means-to-me/, Accessed April 25, 2020

[20] Bass, L., Weber, IO., Zhu, L. (2015) "DevOps: A Software Architect's Perspective", *Addison Wesley*, US.

[21] Dyck A., Penners R., Lichter H. (2015) "Towards Definitions for Release Engineering and DevOps", *IEEE/ACM 3rd International Workshop on Release Engineering*.

[22] Kersten, M. (2018) "What Flows through a Software Value Stream?," *in IEEE Software*, Vol.35-4, pp8-11.

[23] Horlach, B., Drews, P., Schirmer, I., and Boehmann, T. (2017) "Increasing the Agility of IT Delivery: Five Types of Bimodal IT Organization," *Hawaii International Conference on System Sciences*, USA.

[24] The Countries Most (and Least) Likely to be Affected by Automation, https://hbr.org/2017/04/the-countries-most-and-least-likely-to-be-affected-by-automation, Accessed April 18, 2020

[25] World Economic Forum - Strategies for the New Economy Skills as the Currency of the Labour Market, http://www3.weforum.org/docs/WEF_2019_Strategies_for_the_New_Economy_Skills.pdf, Accessed April 18, 2020\

[26] Larjovuori, R.L., Bordi, L., Heikkilä-Tammi, K. (2018) Leadership in the digital business transformation" *Proceedings of the 22nd International Academic Mindtrek Conference (Mindtrek)*, Association for Computing Machinery, USA

[27] Beck K. (2001) "Principles behind the Agile Manifesto", *Agile Alliance*.

[28] Poppendieck, M. and Poppendieck, T. (2003) "Lean Software Development: An Agile Toolkit", *Addison-Wesley Professional*, Boston.

[29] Takeuchi, H. and Nonaka, I. (1986) "The New New Product Development Game", *Harvard Business Review*, Vol. 64, pp137-146.

[30] Sutherland, J., Schwaber, K. (2017) "The Definitive Guide to Scrum: The Rules of the Game", USA.

[31] Levinson, W.A. (2002) "Henry Ford's Lean Vision: Endurng Principles from the First Ford Motor Plant", *Productivity Press*, New York.

[32] Shingo, S., Dillon, A.P. (1988) "A study of the Toyota production system", *Productivity Press*, New York.

[33] Ohno, T. (1988) "Toyota Production System: Beyond Large- Scale Production", *Productivity Press*, New York.

[34] Krafcik, J. F. (1988) "Triumph of the Lean Production System", *Sloan Management Review*, Vol. 30 (1), pp41–52.

[35] Womack, J.P. and Jones, D.T. (1990) "The Machine That Changed the World", *Rawson Associates*, New York.

[36] Poppendieck, M. and Poppendieck, T. (2007) "Implementing Lean Software Development – From Concept to Cash", *Pearson Education*, Boston.

[37] Rodríguez, P., Partanen, J., Kuvaja, P., Oivo, M. (2014) "Combining Lean Thinking and Agile Methods for Software Development A Case Study of a Finnish Provider of Wireless Embedded Systems", *Proceedings of the Annual Hawaii International Conference on System Sciences*, pp4770-4779.

[38] Ravichandran, A., Taylor, K., Waterhouse, P. (2017) "DevOps for Digital Leaders, Reignite Business with a Modern DevOps-Enabled Software Factory", Winchester.

[39] Jabbari R., Ali N.B., Petersen K., Tanveer B. (2016) "What is DevOps? A Systematic Mapping Study on Definitions and Practices", *Proceedings of the Scientific Workshop Proceedings of XP2016*, Edinburgh.

[40] De França, B.B.N., Jeronimo, H., Travassos, G.H. (2016) "Characterizing DevOps by Hearing Multiple Voices", *Proceedings of the 30th Brazilian Symposium on Software Engineering (SBES)*, Association for Computing Machinery, New York, pp53–62.

[41] Lwakatare, Lucy Ellen & Kuvaja, Pasi & Oivo, Markku. (2016) "An Exploratory Study of DevOps: Extending the Dimensions of DevOps with Practices", *The Eleventh International Conference on Software Engineering Advances (ICSEA)*, Rome.

[42] Smeds, J., Nybom, K., Porres, I. (2015) "DevOps: A Definition and Perceived Adoption Impediments", *16th International Conference on Agile Software Development (XP)*, Springer International Publishing, pp. 166–177.

[43] Dingsøyr T., Lassenius C. (2016) "Emerging themes in agile software development: Introduction to the special section on continuous value delivery", *Information and Software Technology*, pp. 56-60.

[44] Lwakatare L.E., Kuvaja P., Oivo M. (2016) "Relationship of DevOps to Agile, Lean and Continuous Deployment", PROFES, pp. 399-415.

[45] Myrbakken H., Colomo-Palacios R. (2017) "DevSecOps: A Multivocal Literature Review", *Software Process Improvement and Capability Determination, SPICE*, Communications in Computer and Information Science, Vol.770, Springer, Cham.

[46] Drews, P., Schirmer, I., Horlach, B., Tekaat, C. (2017) "Bimodal Enterprise Architecture Management: The Emergence of a New EAM Function for a BizDevOps-Based Fast IT", *IEEE 21st International Enterprise Distributed Object Computing Workshop (EDOCW)*, pp57-64.

[47] Kropp, M., Meier, A., Anslow, C., Biddle, R. (2018) "Satisfaction, Practices, and Influences in Agile Software Development", *Proceedings of the 22nd International Conference on Evaluation and Assessment in Software Engineering (EASE)*, Association for Computing Machinery, USA, pp112–121.

[48] Wiedemann, A., Wiesche, M. (2018) "Are You Ready for DevOps? Required Skill Set for DevOps Teams", *Proceedings of the European Conference on Information Systems (ECIS)*, Portsmouth.

[49] Sánchez-Gordón, M., Colomo-Palacios, R. (2018) "Characterizing DevOps Culture: A Systematic Literature Review", *Software Process Improvement and Capability Determination (SPICE)*, Communications in Computer and Information Science, Vol.918, Springer.

[50] Luz, P.W., Pinto, G., Bonifácio, R. (2019) "Adopting DevOps in the real world: A theory, a model, and a case study", *Journal of Systems and Software*, pp157.

[51] Masombuka, T., Mnkandla, E. (2018) "A DevOps collaboration culture acceptance model", *Proceedings of the Annual Conference of the South African Institute of Computer Scientists and Information Technologists (SAICSIT)*, Association for Computing Machinery, USA, pp279–285.

[52] The 4 Roles of DevOps Leadership, https://devopscon.io/blog/the-4-roles-of-devops-leadership, Accessed 26/04/2020

[53] Sahu, S., Pathardikar, A., Kumar, A. (2018) "Transformational leadership and turnover: Mediating effects of employee engagement, employer branding, and psychological attachment", *Leadership & Organization Development Journal*, Vol. 39, pp82-99.

[54] Avolio, B.J., Gardner W.L. (2005) "Authentic leadership development: Getting to the root of positive forms of leadership", *The Leadership Quarterly*, Vol.16-3.

[55] Rudder, C.: 8 habits of successful DevOps team leaders, https://enterprisersproject.com/article/2019/11/devops-habits-successful-leaders, Accessed 26/04/2020

[56] Dubinsky, Y., Hazzan, O (2010) "Ad-hoc leadership in agile software development environments" *Proceedings of the ICSE Workshop on Cooperative and Human Aspects of Software Engineering (CHASE)*, Association for Computing Machinery, USA.

[57] Puppet, DORA, State of DevOps Report 2017, https://puppet.com/resources/report/2017-state-devops-report, Accessed 26/04/2020

[58] Commission Recommendation of 6 May 2003 concerning the definition of micro, small and medium-sized enterprises, https://eur-lex.europa.eu/eli/reco/2003/361/oj, Accessed April 18, 2020

[59] Martin, K., Osterling, M. (2014) "Value Stream Mapping: How to Visualize Work and Align Leadership for Organizational Transformation", *McGraw-Hill Education*, UK.

[60] Mujtaba, S., Feldt, R., Petersen, K. (2010) "Waste and Lead Time Reduction in a Software Product Customization Process with Value Stream Maps", *21st Australian Software Engineering Conference*, pp139-148.

**Authors**

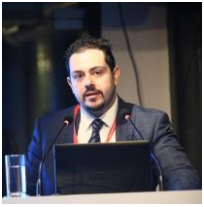

Krikor Maroukian gained a BSc. (Hons) degree in Computer Science and a MSc. in Applied Informatics. Currently, he works as a Sr Modern Service Management consultant at Microsoft for Central Eastern Europe. He is also the International Representative of itSMF Hellas and BoD member of the British Computer Society (BCS), The Chartered Institute for IT, Hellenic Section. In the past, he has held Project Management and Security Management positions. Krikor is a PhD student at Henley Business School, University of Reading, UK. Krikor's interest lie in the areas of service management, agile software development, lean product development, DevOps adoption. Krikor is author of Axelos ITIL4® – Managing Professional High Velocity IT. Krikor was awarded the Highly Commended Paper in the 2017 Emerald Literati Network Awards for Excellence.

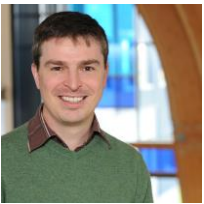

Stephen R. Gulliver received a BEng. (Hons) degree in Microelectronics, an MSc. degree (Distributed Information Systems) and a PhD in 1999, 2001, and 2004 respectively. Stephen worked within the Human Factors Integration Defence Technology Centre (HFI DTC), before getting a job as a lecturer at Brunel University (2005-2008). Dr Gulliver joined Henley Business School (University of Reading) in 2008 as a lecturer and in 2014 was promoted to the role of Associate Professor. Since 2005 Dr Gulliver's teaching and research (in the UK and abroad) has linked to the area of pervasive Informatics, and he has interests including: multimedia and information assimilation, e-learning and education systems, usability and human factors, technology acceptance, persuasion systems, health systems, and systems conflict and failure.